\documentclass[conference]{IEEEtran}
\IEEEoverridecommandlockouts

\usepackage{cite}
\usepackage{amsmath,amssymb,amsfonts}
\usepackage{algorithmic}
\usepackage{graphicx}
\usepackage{textcomp}
\usepackage{xcolor}

\usepackage{graphicx}
\usepackage{enumitem}
\usepackage{booktabs,ragged2e}
\usepackage[flushleft]{threeparttable}
\usepackage{colortbl}
\usepackage{booktabs}
\usepackage{tabularx}
\usepackage{multirow}
\usepackage{hyperref}

\def\BibTeX{{\rm B\kern-.05em{\sc i\kern-.025em b}\kern-.08em
    T\kern-.1667em\lower.7ex\hbox{E}\kern-.125emX}}
\begin{document}

\title{Distance Based Single-Channel \\ Target Speech Extraction\\
}

\author{\IEEEauthorblockN{1\textsuperscript{st} Runwu Shi}
\IEEEauthorblockA{\textit{Dept. of Systems \& Control Engineering} \\
\textit{Tokyo Institute of Technology}\\
Tokyo, Japan \\
shirunwu@ra.sc.e.titech.ac.jp}
\and
\IEEEauthorblockN{2\textsuperscript{nd} Benjamin Yen}
\IEEEauthorblockA{\textit{Dept. of Systems \& Control Engineering} \\
\textit{Tokyo Institute of Technology}\\
Tokyo, Japan \\
benjamin.yen@ieee.org}
\and
\IEEEauthorblockN{3\textsuperscript{rd} Kazuhiro Nakadai}
\IEEEauthorblockA{\textit{Dept. of Systems \& Control Engineering} \\
\textit{Tokyo Institute of Technology}\\
Tokyo, Japan \\
nakadai@ra.sc.e.titech.ac.jp}

%
}

\maketitle

\begin{abstract}
This paper aims to achieve single-channel target speech extraction (TSE) in enclosures by solely utilizing distance information. This is the first work that utilizes only distance cues without using speaker physiological information for single-channel TSE. Inspired by recent single-channel Distance-based separation and extraction methods, we introduce a novel model that efficiently fuses distance information with time-frequency (TF) bins for TSE. Experimental results in both single-room and multi-room scenarios demonstrate the feasibility and effectiveness of our approach. This method can also be employed to estimate the distances of different speakers in mixed speech. Online demos are available at \url{https://runwushi.github.io/distance-demo-page/}.
\end{abstract}


\section{Introduction}
Target Speech Extraction (TSE) is the task that utilizes speaker-relevant clues to extract target speech from an audio mixture consisting of multiple speakers. Such speaker clues include enrolled voice, facial images, and lip movement \cite{Zmolikova-NeuralTargetSpeechExtraction-2023, ijcai2024p709}. However, in real-world scenarios, ideal speaker-related information may not always be readily available, and such sensitive data that can be used for biometrics may raise privacy concerns \cite{noe2020adversarial}. Previously, several studies have reported successful results using spatial and distance clues for speech separation and TSE, which do not explicitly import speaker-relevant information, but instead utilize the transfer properties between the speaker and the microphone in enclosed rooms. The distance clue works mainly because the direct-to-reverberation ratio (DRR) decreases as the target to microphone distance increases within the enclosures \cite{Kushwaha-SoundSourceDistanceEstimation-2023, Patterson-DistanceBasedSoundSeparation-2022}. The late reverberation component is less affected by distance, while the direct and the early components are inversely proportional to the distance. \\
The first exploration of using distance clue for deep learning-based single-channel speech separation or extraction is Distance Based Sound Separation (DSS) \cite{Patterson-DistanceBasedSoundSeparation-2022}, in which a basic recurrent neural network (RNN) model is used to separate the single-channel mixed audio into two groups: near and far group according to a static distance threshold such as 1.5m. This model implicitly learns the acoustic properties of mixed reverberated speech located in the near and far positions and separates them. However, due to the static threshold, users have to retrain the model for a new threshold. \\
To make such threshold-based methods more flexible, \cite{gu2024rezero}  proposes the Region-customizable Sound Extraction. It uses a dynamic distance threshold to directly control the boundary between the near and far groups. The target distance threshold value is converted into a learnable embedding and is then fused with the multi-channel signal intermediate representations to obtain the target near sound group. However, when the expected source is located at a specific distance range, the model needs to infer multiple times for the target sound. \\
Following the DSS, \cite{Lin-FocusSoundYouMonaural-2023} also adopts the idea of static distance threshold but realizes in the monaural TSE manner where the model only outputs one target speech. In detail, the target speaker embedding is learned by the network’s speaker encoder, and this encoder will only output the embedding of the speaker within the distance threshold. This work only tests scenarios involving a single speaker within the threshold range, and this method, like \cite{Patterson-DistanceBasedSoundSeparation-2022}, is also limited to a static threshold. Moreover, during the training process, speaker information is necessary to classify the learned speaker embeddings, which reduces the emphasis on distance information. The distance is primarily used to train the model to produce the desired near-speaker embedding, and the generation of the target speech depends solely on the obtained speaker embedding, rather than the distance information. In summary, only a few studies have considered distance cues \cite{taherian2022multi,petermann2024hyperbolic}, and there remains significant room for exploration in this area.\\
Rather than employing any speaker embedding to extract target speech, in this work, we propose Distance-based Single-Channel Target Speech Extraction, in which we rely solely on distance information as the cue for TSE, without any prior physiological information about the speaker. Our approach leverages the spatial reverberation cues inherent in an enclosed room, where the distance from the microphone varies among different speakers. This allows us to utilize the raw distance information to extract speech at specific distances. Furthermore, distance serves as an ambiguous cue, as identical distances can correspond to multiple room impulse responses (RIR) from different directions. This ambiguity renders Distance-based TSE feasible in a single-channel manner, which forms the basis of our proposed method. To our knowledge, this work is the first to propose utilizing only distance information for single-channel TSE in enclosures. For this task, we introduce a model that efficiently fuses the distance information and the TF bins, and the results demonstrate its feasibility. The remainder of this paper is organized as follows: Section 2 describes the proposed method. Section 3 presents the dataset and experimental results, and Section 4 concludes the paper. 
\section{Methodology}
\subsection{Problem formulation}
\begin{figure}[!t]
\begin{minipage}[b]{1.0\linewidth}
  \centering
  \centerline{\includegraphics[width=8.5cm]{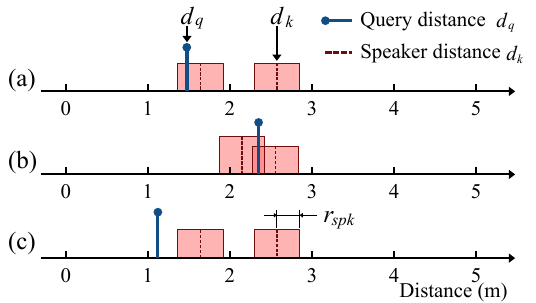}}
\end{minipage}
\caption{Query distance $d_q$ and speaker distance $d_k$.}
\label{fig:one}
\end{figure}

Assuming $K$ speakers in an enclosed room, and denote $s_{k}(t)$, $x_k(t)$ and $h_k(t)$ as the $k$ th speaker's anechoic speech, reverberant speech, and corresponding RIR, respectively. The reverberant speech $x_k(t)$ can be formulated as 
\begin{align}
  x_k(t) = s_{k}(t) \, \ast \, h_k(t),
  \label{equation:first}
\end{align}
where ${\ast}$ represents the convolution operator in the temporal domain. For a single-channel microphone in the room. the collected mixture signal can be represented as 
\begin{align}
  y = \sum_{k=1}^{K} x_k(t).
  \label{equation:second}
\end{align}
Each target speech $x_k(t)$ corresponds to an impulse response $h_k(t)$ which is influenced by the room's physical features and the positions of the microphone and the speaker, and the positions are represented by their relative distance $d_k$. Given a query distance $d_q$, the TSE model will output the target speech located in the range $|d_k-d_q| \leq r_{spk}$, in which the $r_{spk}$ represents the speaker distance range centered with speaker distance, as shown in Fig~\ref{fig:one} (a), which can be depicted as
\begin{align}
y \xrightarrow{d_q} \sum_{k} x_k(t), 
k \text{ s.t. } |d_k - d_q| \leq r_{spk}.
\end{align}
Speakers may have similar distances, and when the query distance $d_q$ lies within the range of more than one speaker, the model should output the sum of speeches, as shown in Fig~\ref{fig:one} (b). In the case where no speaker exists near the query distance, the model outputs zero, as shown in Fig~\ref{fig:one} (c), depicted as
\begin{align}
y \xrightarrow{d_q} 0, 
\forall k, \ |d_k - d_q| > r_{spk}.
\end{align}
\subsection{Proposed model}
\begin{figure}[!tb]
\begin{minipage}[b]{1.0\linewidth}
  \centering
  \centerline{\includegraphics[width=8.5cm]{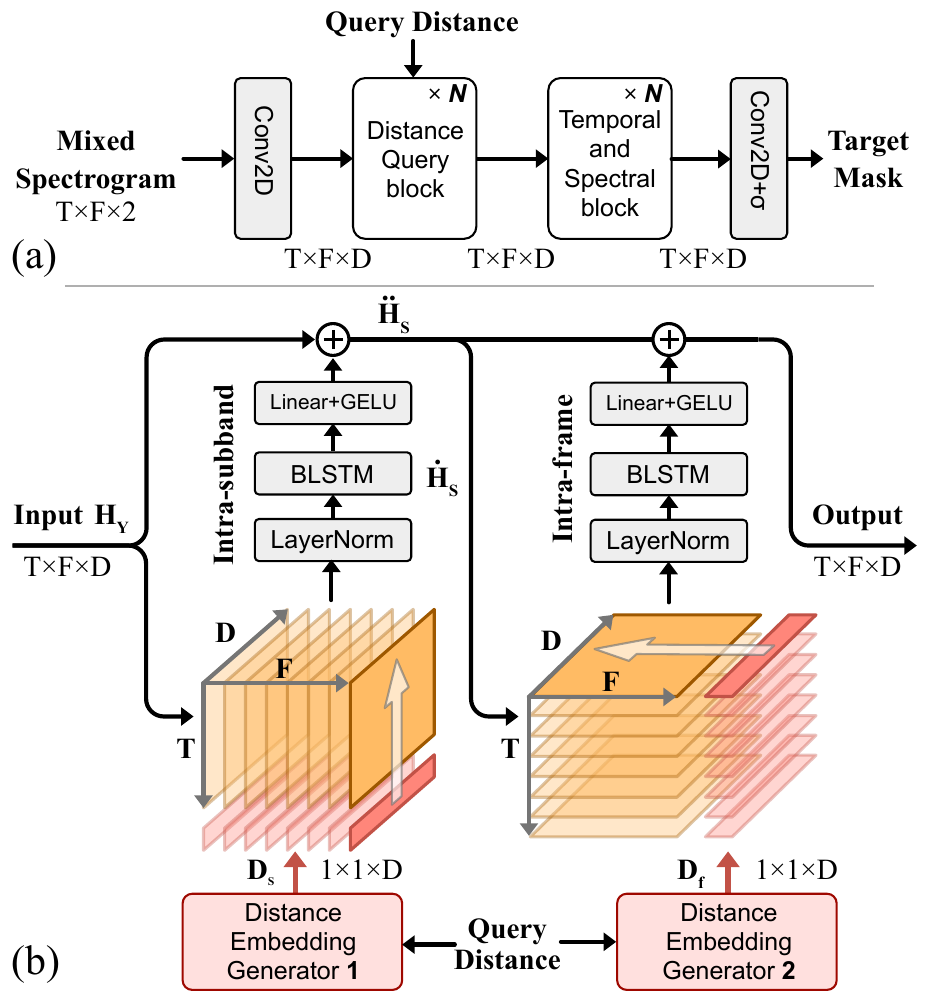}}
\end{minipage}
\caption{(a) Overall structure, (b) Structure of Distance Query block.}
\label{fig:two}
\end{figure}

TSE's core problem is identifying the target speaker given the clue, so the general TSE framework always consists of a clue encoder and a speech extraction module. The clue encoder is designed to project the clue to the target speaker embedding, and the speech extraction module is mainly used to fuse the obtained embedding with the representation of the mixed speech \cite{Zmolikova-NeuralTargetSpeechExtraction-2023}. Our method follows this framework in the TF domain. The overall structure is shown in Fig~\ref{fig:two} (a). 
\vspace{0pt} 
First, by converting $y(t)$ into the TF domain via short-time Fourier Transform (STFT), the input feature of the model is the concatenated real and imaginary (RI) components $ \mathbf{Y} \in \mathbb{R}^{2 \times T \times F} $, which is then processed by the encoder consisting of the initial 2D convolution layer with $3 \times 3$ kernel and the global layer normalization to obtain the $D$-dimensional TF embeddings $ \mathbf{H_Y} \in \mathbb{R}^{D  \times T \times F} $ \cite{Wang-TFGridNetIntegratingFullSubBand-2023}. The TF embeddings $ \mathbf{H_Y}$ are then fed into the Distance Query (DQ) block and Temporal and Spectral (TS) block stacks. Finally, the decoder containing a 2D convolution layer with $3 \times 3$ kernel and a Sigmoid function outputs the target speech mask. The DQ block contains learnable distance embeddings that fuse with the intermediate representation in both temporal and spectral domains, and the TS block is designed for outputting target speech. These two kinds of blocks share the same basic structure and will be introduced in this section.

\subsubsection{Distance Query block}
The main function of the DQ module is to receive the target query distance and TF embedding $ \mathbf{H_Y}$ and output the intermediate representation of the target speech. The specific structure is shown in Fig~\ref{fig:two} (b), which contains two learnable distance embeddings, an intra-subband fusion module, and an intra-frame fusion module. 
\\
The learnable distance embedding generator (DEG) contains three linear layers followed by the tanh function. It receives the original query distance $d_q$ as input, and outputs distance embedding $ \mathbf{D} \in \mathbb{R}^{1 \times D} $,  which is similar to \cite{gu2024rezero} except that the last layer has no activation function. The intra-subband fusion module and the intra-frame fusion module are each associated with a corresponding DEG, which outputs $\mathbf{D_s}$ and $\mathbf{D_f}$, respectively. For the intra-subband fusion module, it fuses the TF embedding $\mathbf{H_Y}$ and distance embeddings $\mathbf{D_s}$ within a single subband using a single bidirectional long short-term memory (BLSTM) layer. This BLSTM layer is generalized across subbands and can be viewed as retrieving target distances in different frames of all subbands, which provides an efficient way to save parameters \cite{Wang-DasformerDeepAlternatingSpectrogram-2023a, Wang-TFGridNetIntegratingFullSubBand-2023, Luo-DualpathRNNEfficientLong-2020}. In detail, the input TF embedding $\mathbf{H_Y}$ is firstly concatenated with the distance embeddings $\mathbf{D_s}$ that is replicated along the spectral dimension. The concatenated $\mathbf{H_s}$ is then processed by the Layer Normalization (LN) and the BLSTM layer,
\begin{align}
  \mathbf{H_s} = \text{Concat}(\mathbf{H_Y}, \mathbf{D_s}) \in \mathbb{R}^{D \times (T+1) \times F},
  \label{equation:four}
\end{align}
\begin{align}
  \mathbf{\dot{H}_s} &= [\text{BLSTM}(\text{LN}{(\mathbf{H_s}[:,t,:])}), \text{for } t = 1,...,(T+1)] \notag \\
  &\in \mathbb{R}^{H \times (T+1) \times F},
  \label{equation:five}
\end{align}
\noindent
where the $H$ represents the dimension of hidden states. Finally, the hidden states of each TF embedding are fed into one linear layer with a GELU activation function to project to the original shape and connected using residuals,
\begin{align}
  \resizebox{0.9\linewidth}{!}{$
    \mathbf{\ddot{H}_s} = \mathbf{H_Y}+\text{GELU}(\text{Linear}(\mathbf{\dot{H}_s}[:,0:T,:])) \in \mathbb{R}^{D \times T \times F}.
  $}
  \label{equation:six}
\end{align}
The intra-frame fusion module shares the same structure as the intra-subband fusion module, except that the concatenation with distance embedding $\mathbf{D_f}$ is along the spectral dimension, and the scanning direction of BLSTM is across subbands in each frame. 

\subsubsection{Temporal and Spectral block}
After fusing the target distance embedding and the speech representation, the TS block is utilized to enhance the quality of the output speech. The TS block shares the basic structure with the DQ block without the fusing part, which repetitively utilizes one RNN across all frames or subbands \cite{Luo-DualpathRNNEfficientLong-2020}. 

\subsection{Loss function}

Two kinds of loss functions are used for the model to output both target speech and zero value conditioned on the target distance. Given the target speech $\mathbf{x}_s$, the estimated speech $\hat{\mathbf{x}_s}$, we adopt the signal-to-distortion ratio (SDR) as the loss function when the speakers exist at the query distance, 

\begin{align}
  \scalebox{1.0}{$
    L_{active}(\mathbf{x}_s, \hat{\mathbf{x}_s}) = 10 \log_{10} \left( \frac{\|\mathbf{x}_s\|^2}{\| \mathbf{x}_s - \hat{\mathbf{x}_s} \|^2 + \tau \|\mathbf{x}_s\|^2} \right)
  $},
\end{align}
\noindent
where the $\tau$ is the soft threshold to control the upper bound of the loss $\tau = 10^{- \frac{\eta}{10}}$, in which $\eta$ is set to 30 dB. The widely used scale invariant SDR (SI-SDR) \cite{Roux-SDRHalfbakedWellDone-2019} is not considered since we wish the model could handle inactive source conditions that the model could output zero. In our experiments, we find that SI-SDR typically causes the model to output small scales under all conditions (presence of target speech and zero output), whereas in tasks that need to detect active or inactive sources, correctly constructing signal scale is crucial \cite{Delcroix-SoundBeamTargetSoundExtraction-2022}. For the condition that no speaker is near the query distance, the adopted loss function is the inactive SDR (iSDR) \cite{Wisdom-WhatAllFussFree-2021, Delcroix-SoundBeamTargetSoundExtraction-2022},

\begin{align}
  L_{inactive}(\mathbf{y}, \hat{\mathbf{x}_s}) = 10 \log_{10} \left({\| \hat{\mathbf{x}_s}\|^2 + \tau^{inactive} \| \mathbf{y} \|^2} \right),
\end{align}
\noindent
where $\mathbf{y}$ is the mix speech, and the $\tau^{inactive}$ represents the soft threshold which is set to $10^{-2}$.

\section{Experiment}
\label{sec:pagestyle}

\subsection{Dataset generation}
In this work, the mixed speech input is the superposition of two reverberant speech signals, each convolved with a position-specific RIR, and we utilize both simulated and realistic RIRs. For the simulated dataset, we use the randomized image method \cite{Scheibler-PyroomacousticsPythonPackageAudio-2018}. For the realistic dataset, we adopt the BUT ReverbDB dataset \cite{Szoke-BuildingEvaluationRealRoom-2019}, where we use three rooms with varying sizes. In total, we construct four datasets: three with simulated RIRs and one with real RIRs.
\subsubsection{\textbf{D1} Fixed room size and fixed microphone position}
This simulated dataset contains a fixed room with the size of $7 \times 8 \times 3$ meters and a fixed microphone position at $3.5 \times 4 \times 1.1$ meters. The reverberation time (RT60) is set to 0.2 seconds. Speaker positions are randomly initialized at least 0.5 meters from the walls, with heights ranging from 1.2 to 2.0 meters. A total of 10,000 RIRs are generated. The dataset is split into training, validation, and testing datasets with a ratio of 0.9:0.02:0.08.
\subsubsection{\textbf{D2} Fixed room size and random microphone positions}
This simulated dataset has a fixed room with the size of $7 \times 8 \times 3$ meters. We randomly generate 100 microphone positions, each corresponding to 2,000 randomly generated speaker positions, resulting in 200,000 RIRs. The dataset is split into training, validation, and testing datasets with a ratio of 0.9:0.01:0.09.
\subsubsection{\textbf{D3} Random room sizes and random microphone positions}
This simulated dataset has 50,000 rooms randomly initialized within the range from $4 \times 5 \times 2.5$ meters to $8 \times 10 \times 3.0$ meters. The RT60 is randomly initialized from 0.2s to 0.5s. Each room contains one randomly placed microphone and ten speakers, resulting in 489,813 RIRs, since smaller sized rooms may not have longer distance samples. Training, validation, and test datasets are divided based on different rooms with a ratio of 0.9:0.02:0.08. 
\subsubsection{\textbf{D4} Real RIRs from BUT ReverbDB dataset}
We adopt the real RIRs collected from the three rooms Q301, L207, and L212 in the dataset. The distance ranges from 0.266 to 10.521 meters. A total of 624 RIRs are used for fine-tuning, and 182 RIRs are reserved for testing.\\
The speech is adopted from LibriLight \cite{Kahn-LibriLightBenchmarkASRLimited-2019} dataset, and 128, 48, and 64 different speakers are used for training, validation, and testing. Each utterance is convolved from randomly chosen speech and RIR, and the root mean square energy of the convolved signal is randomly adjusted from -25 to -20 dB. 
\subsection{Training process and configuration}
The model contains four DQ blocks and four TS blocks, the dimension size $D$ is set to 64, and the hidden state size $H$ is set to 64. The three linear layers in each distance embedding generator contain 32, 64, and 64 units, respectively. The parameter amount is 1.25M due to the efficient structure. A frame length of 32 ms with 16 ms frame shift is used in STFT. The $r_{spk}$ is set to 0.5 meters for D1 to D3 and to 0.1 meters for D4. In the training stage, we use the Adam \cite{Kingma-AdamMethodStochasticOptimization-2017} optimizer, and gradient clipping is set to a maximum norm of 5. Batch size is set to 14. The initial learning rate is 0.001, and if no lower loss is found for 14 consecutive epochs, the learning rate reduces to 80\%. The training has 500 epochs, and the ratio between the presence and absence of the target speech is set to 0.9:0.1 for the first 250 epochs and 0.7:0.3 for the last 250 epochs. For scenario D4, the model is fine-tuned using the model pre-trained on D3 due to data scarcity.

\subsection{Experiment results}
\subsubsection{Results in different scenarios}
The evaluation needs to consider both the presence and absence of speakers at the query distance. For the presence condition, we use SDR, SDR improvement (SDRi), and PESQ to verify speech performance and scale accuracy. For the absence condition, we use the iSDR as the metric. The evaluation results on two speaker mixtures are presented in Table~\ref{tab:one}. The results are obtained by repeating testing 5 times and expressed as mean $\pm$ standard deviation.


\setlength{\textfloatsep}{5pt plus 1.0pt minus 1.0pt}

\begin{table}[!t]
  \caption{Results of two speaker mixture under four conditions.}
  \label{tab:one}
  \centering
  \renewcommand{\arraystretch}{1.25} 
  \begin{threeparttable}
  \begin{tabularx}{\columnwidth}{>{\arraybackslash}b{1.25cm} >{\centering\arraybackslash}b{1.5cm} >{\centering\arraybackslash}b{1.5cm} >{\centering\arraybackslash}X >{\centering\arraybackslash}X >{\centering\arraybackslash}X}  
  \specialrule{1.2pt}{1pt}{1pt}
    Dataset & SDR$\uparrow$ (dB) & SDRi$\uparrow$ (dB) & PESQ$\uparrow$ & iSDR$\downarrow$ \\ 
  \midrule 
    \textbf{D1} 1r\&1m & \text{$12.47 \pm 0.12$}  & \text{$8.83 \pm 0.24$} & \text{$2.40 \pm 0.02$} & \text{$9.59 \pm 0.08$}\\
    \textbf{D2} 1r\&ms & \text{$10.43 \pm 0.17$} & \text{$7.29 \pm 0.42$} & \text{$2.20 \pm 0.02$} & \text{$13.85 \pm 0.07$}\\
    \textbf{D3} rs\&ms & \text{$6.83 \pm 0.13$} & \text{$3.36 \pm 0.12$} & \text{$1.75 \pm 0.003$} & \text{$18.46 \pm 0.55$} \\
    \textbf{D4} Real & \text{$4.79 \pm 0.11$} & \text{$3.68 \pm 0.14$} & \text{$1.67 \pm 0.02$} & \text{$36.76 \pm 0.46$} \\
  \specialrule{1.2pt}{1pt}{1pt}
  \end{tabularx}
  \end{threeparttable}
\end{table}

\setlength{\textfloatsep}{5pt plus 1.0pt minus 1.0pt}

\begin{table}[!t]
  \caption{Comparison results of different models trained on D1.}
  \label{tab:2}
  \centering
  \renewcommand{\arraystretch}{1.25} 
  \begin{threeparttable}
  \begin{tabularx}{\columnwidth}{>{\centering\arraybackslash}b{1.2cm} >{\centering\arraybackslash}X >{\centering\arraybackslash}X >{\centering\arraybackslash}X >{\centering\arraybackslash}X}  
  \specialrule{1.2pt}{1pt}{1pt}
    Model & SDR$\uparrow$ (dB) & SDRi$\uparrow$ (dB) & PESQ$\uparrow$ & iSDR$\downarrow$ \\ 
  \midrule 
    \multirow{1}{*}{LSTM+DEG} & \text{$4.31 \pm 0.18$} & \text{$1.13 \pm 0.16$} & \text{$1.56 \pm 0.02$} & \text{$14.50 \pm 0.19$} \\
    \multirow{1}{*}{2DQ+6TS} &  \text{$11.45 \pm 0.24$}& \text{$8.11 \pm 0.23$} & \text{$2.10 \pm 0.03$} & \text{$9.45 \pm 0.19$} \\
    \multirow{1}{*}{6DQ+2TS} &  \text{$12.11 \pm 0.27$} & \text{$8.63 \pm 0.20$} & \text{$2.31 \pm 0.03$} & \text{$8.65 \pm 0.15$} \\
  \specialrule{1.2pt}{1pt}{1pt}
  \end{tabularx} 
  \end{threeparttable}
\end{table}

In the four test conditions, the proposed method achieves an SDR of 12.47 dB in the fixed room and microphone scenario. Performance degrades as the number of rooms and microphones increases. The diverse room sizes and random microphone placements pose a significant challenge, as the method must handle numerous combinations of room and microphone positions for a given query distance. The results on the real dataset are lower, primarily due to the multi-room conditions and limited data availability. A more extensive dataset is necessary for further improvement.

\subsubsection{Comparison results}
We compare the proposed method with the basic LSTM model similar to DSS \cite{Patterson-DistanceBasedSoundSeparation-2022}, which has 5.98M parameters and contains 4 Bi-LSTM layers and a final linear layer with Sigmoid function for mask prediction. We keep the DEG and concatenate the distance embedding with the mixture spectrogram for TSE. We also adjusted the number of DQ and TS blocks in the proposed method. The results in Table~\ref{tab:2} demonstrate the performance of the proposed method.



\subsubsection{Speaker distance estimation in mixed speech}
\begin{figure}[!t]
\begin{minipage}[b]{1.0\linewidth}
  \centering
  \hspace*{-0.5cm} 
  \centerline{\includegraphics[width=8.5cm]{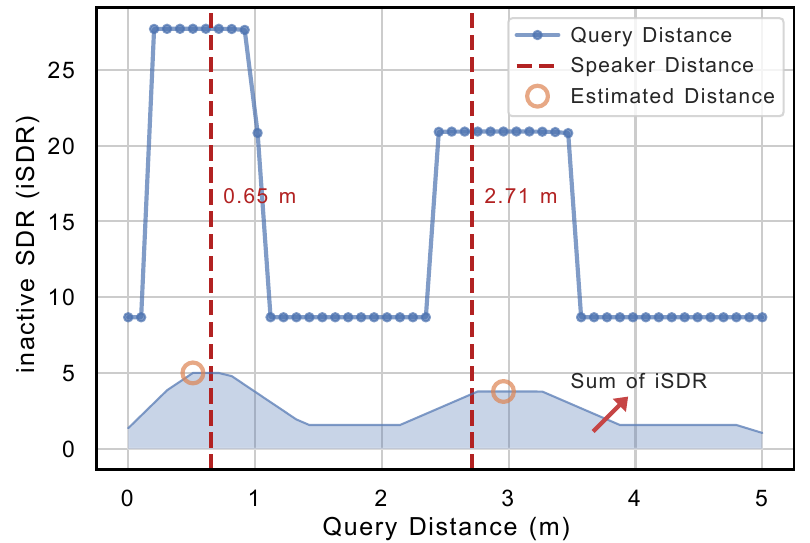}}
\end{minipage}
\caption{Inactive SDR of speech versus different query distance. The red dashed line represents the ground truth speaker distance and the blue point represents the query distance, higher iSDR is an indication of the speaker's presence.}
\label{fig:four}
\end{figure}

One notable advantage of the proposed Distance-based TSE model is its ability to output speech within the range centered around the query distance. This feature enables the detection of speech at varying distances by iteratively processing all possible input distances. We use the sum of the iSDRs of neighboring distances as the indicator. For instance, we utilize the model trained on scenario D1 and set the query distance interval to 0.5m. The results are illustrated in Fig~\ref{fig:four}. The blue shaded area is the iSDR sum within 1m and the orange circle is the detection peak value meaning the detected speaker distance. Table~\ref{tab:three} presents the mean absolute error (MAE), quantifying the difference between the highest detection peak and the closest real speaker distance.

\setlength{\textfloatsep}{5pt plus 1.0pt minus 1.0pt}

\begin{table}[!t]
  \caption{Distance estimation results.}
  \label{tab:three}
  \centering
  \begin{threeparttable}
  \begin{tabularx}{0.8\columnwidth}{>{\arraybackslash}p{1.5cm} 
  >{\centering\arraybackslash}p{1.5cm} >{\centering\arraybackslash}p{3cm}}  
  \specialrule{1.2pt}{1pt}{1pt}
    Dataset & MAE (m) $\downarrow$ & \text{Estimated range (m)} \\ 
  \midrule 
    \multirow{1}{*}{\textbf{D1} 1r\&1m} & 0.26 & 0 - 5\\
    \multirow{1}{*}{\textbf{D2} 1r\&ms} & 0.32 & 0 - 5\\
    \multirow{1}{*}{\textbf{D3} rs\&ms} & 0.41 & 0 - 5\\
    \multirow{1}{*}{\textbf{D4} Real} & 1.09 & 0 - 10\\
  \specialrule{1.2pt}{1pt}{1pt}
  \end{tabularx}
  \end{threeparttable}
\end{table}

\section{Conclusion}
We propose an innovative task: Distance-based single-channel target speech extraction and design a parameter-efficient model to achieve it. The results demonstrate the task's feasibility and potential, including the ability to estimate speaker distance from the mixture. Building on these findings, our future work will focus on enhancing performance in real scenarios and multi-speaker situations.


\bibliographystyle{IEEEtran}
\bibliography{strings}

\vspace{12pt}

\end{document}